\documentclass{jpp}
\usepackage{graphicx}
\usepackage{epstopdf, epsfig, color}
\usepackage{amsmath, amssymb, bm}

\newcommand{\Tpush}{\mathsf{T}_{s\textrm{gy}}^{-1}}
\newcommand{\Tpushbar}{\mathsf{\bar{T}}_{\bar{s}\textrm{gy}}^{-1}}
\newcommand{\Tpull}{\mathsf{T}_{s\textrm{gy}}}

\shorttitle{Differential formulation of the gyrokinetic Landau operator}
\shortauthor{E. Hirvijoki, A. J. Brizard and D. Pfefferl\'e}

\title{Differential formulation of the gyrokinetic Landau operator}

\author{Eero Hirvijoki\aff{1}
  \corresp{\email{ehirvijo@pppl.gov}},
  Alain  J. Brizard\aff{2}
 \and David Pfefferl\'e\aff{1}}

\affiliation{\aff{1}Princeton Plasma Physics Laboratory, Princeton, New Jersey 08543, USA
\aff{2}Department of Physics, Saint Michael's College, Colchester, Vermont 05439, USA}

\begin{document}

\maketitle

\begin{abstract}
  Subsequent to the recent rigorous derivation of an energetically
  consistent gyrokinetic collision operator in the so called Landau
  representation, this paper investigates the possibility of finding a
  differential formulation of the gyrokinetic Landau collision
  operator. It is observed that, while a differential formulation is
  possible in the gyrokinetic phase-space, reduction of the
  resulting system of partial differential equations to 5D via gyroaveraging poses a challenge. Based on the
  present work, it is likely that the gyrocentre analogs of the
  Rosenbluth-MacDonald-Judd potential functions must be kept gyroangle dependent. 
\end{abstract}

\section{Introduction}
\label{sec:intro}
Phase-space Lagrangian perturbation theory, formulated in terms of Lie-transformations and first applied to the guiding-centre motion by \citet{Littlejohn_1982}, provides a rigorous basis for gyrokinetics, in that it yields exact energy and momentum invariants against which gyrokinetic codes can be validated. For a comprehensive overview of the topic, an exhaustive list of references could be pointed out. Instead, to maintain focus, we refer the reader to the review papers by \citet{Brizard_2007} and \citet{Cary_2009} and the references therein. 

In spite of the success of Lie-transform perturbation theory, the current state-of-the-art gyrokinetics tends to deviate from the path of strict rigorousness when discussion turns into the collision operator. It is our purpose to criticize neither the existing theory nor codes, but merely to point out how difficult consistency is to achieve if collisional effects are considered. And by rigour, we refer to the treatment of Finite-Larmor-Radius (FLR) effects that result in the collision operator from the coupling of configuration and velocity space coordinates in the gyrocentre transformation \citep[see, e.g.,][]{Catto_1977,Xu_1991}.

This is not to say that significant effort would not have been made in order to develop collision operators for gyrokinetics. In contrast, several papers \citep[see, e.g.,][]{Abel_2008,Sugama_2009,Li_2011} provide expressions for both model and full linearized collision operators that are suitable for numerical implementation. These linear operators are typically presented in the form of 
\begin{equation*}
C^{\mathrm{GK}}[h]=\left\langle e^{\bm{\rho}\cdot\nabla}C[e^{-\bm{\rho}\cdot\nabla}h]\right\rangle,
\end{equation*} 
where $\bm{\rho}$ is the Larmor radius and $h$ is the nonadiabatic part of the perturbed distribution function. The prevalence of this expression, even during the current era of modern gyrokinetics, probably stems from history: in local flux tube simulations with periodic boundary conditions for the configuration space, Fourier transformation of the above expression admits straight-forward and explicit gyroaveraging if $\bm{\rho}$ is taken to be of zeroth order with respect to the guiding-centre transformation. The caveat in these operators is, however, the lack of modern day push-forward and pull-back operators. While the push-forward--pull-back formalism is rigorously used on the left hand side of the kinetic equation, equally accurate treatment of the collision operator has been all but neglected. After all, using push-forward and pull-back operators, the expression for the linearized collision operator should read
\begin{equation*}
C^{\mathrm{GK}}[h]=\left\langle \mathsf{T}_{\textrm{gy}}^{-1} C[\mathsf{T}_{\textrm{gy}}h]\right\rangle,
\end{equation*} 
where $\mathsf{T}_{\textrm{gy}}^{-1}$ and $\mathsf{T}_{\textrm{gy}}$ denote the push-forward and pull-back operators between the particle and gyrocentre phase-space. We believe the reason behind this discrepancy is the difficulty in explicitly and consistently applying the push-forward to velocity derivatives present in the collision operator. 

Some have derived collisional gyrokinetics even starting directly from many body gyrokinetics \citep[see the elegant paper by][]{Mishchenko_2007}, but the pioneering step towards developing a collision operator, consistent with the modern day gyrokinetics, was put forward by \citet{Brizard_2004}. He understood that the particle phase-space velocity derivative could be presented in terms of the noncanonical particle phase-space Poisson bracket. This discovery lead to the first consistent guiding-centre test particle operator and also to consistent bounce-averaged formalism \citep[see][]{Brizard_2009}. Recently, the development of consistent operators has accelerated \citep[see][]{Madsen_2013} and culminated to the results presented in \citet{Burby_2015}. By this date, the full nonlinear Landau collision operator has been transformed to gyrokinetic phase-space and exact conservation laws together with H-theorem demonstrated up to arbitrary order with respect to the asymptotic gyrocentre transformation. For the first time, the equations behind gyrokinetics can thus be considered whole in the sense that both the Vlasov part and the collision operator can be transformed to the gyrocentre phase-space using the same push-forward and pull-pack operators.

The rigorousness, however, comes with a heavy price tag. The gyrocentre Landau operator is an integro-differential operator, whose evaluation involves a rather complicated 6D integral over the gyrocentre phase-space. In search for salvation, one remembers that, in particle phase-space, the Landau operator can be converted into a system of coupled partial differential equations via the Rosenbluth-MacDonald-Judd (hereafter RMJ) potential functions \citep{Rosenbluth_1957}. One could thus expect the same to be true for the gyrocentre collision operator: after all, the gyrokinetic Landau operator results from the Lie-transformation of the particle phase-space Landau operator. Furthermore, in particle phase-space, the differential representation has allowed the use of fast elliptic solvers to speed up the evaluation of collisions, reducing the computational complexity from ${\cal O}(n^2)$ for the integro-differential to ${\cal O}(n \log n)$ for the differential formulation, as was demonstrated in \citet{Pataki_2011}. To be more precise, the RMJ potentials, in terms of which the velocity space Fokker-Planck coefficients are often expressed, can be computed either by solving the corresponding Poisson equations, a problem that leads to sparse matrix inversion, or by evaluating the corresponding Green's function solutions, a problem which translates to full matrix--vector multiplication. Having similar option for the gyrocentre collision operator could be especially important since the coupling of configuration and velocity space coordinates in gyrokinetics actually increases the dimensionality of the collision operator, in contrast to what is often thought.

The purpose of this paper is, first of all, to present the differential formulation of the gyrocentre Landau collision operator and, secondly, to discuss the nontrivial issues that arise with it. We start by expressing the particle phase-space collisional kinetic theory in terms of the non-canonical Poisson bracket and illustrate the close analogy between using either the bracket or curvilinear velocity space coordinates. Once this connection is demonstrated for the Reader to feel less alienated with expressing everything in terms of brackets, the gyrocentre coordinate transformation is applied to particle phase-space collision operator and the differential form of the gyrokinetic collision operator presented and briefly compared to the Landau version. Finally, we discuss why the reduction of the 6D differential version of the gyrocentre collision operator to a 5D gyroaveraged version poses a challenge: the gyrocentre versions of the RMJ potentials cannot easily be reduced to 5D except in the limit of zero Larmor radius, in which case the operator naturally collapses to the axially symmetric particle phase-space collision operator.

\section{Particle phase-space collision operator}
Let us start by briefly recalling the collisional theory in particle phase-space. In a multispecies plasma, the kinetic equation that describes the evolution of the particle distribution function $f_s$ of species $s$ due to both Hamiltonian motion and collisions with species $\bar{s}$ is written as
\begin{equation}
\frac{\partial f_s}{\partial t}+\dot{z}^{\alpha}\frac{\partial f_s}{\partial z^{\alpha}} =C_{s\bar{s}}[f_s,f_{\bar{s}}]
\end{equation}
where $\bm{z}=(\bm{x},\bm{v})$ are the phase space coordinates, and the Fokker-Planck collision operator is defined
\begin{equation}
C_{s\bar{s}}[f_s,f_{\bar{s}}]\equiv\frac{\partial}{\partial v^i}\left(D_{s\bar{s}}^{ij}\frac{\partial f_s}{\partial v^j}-K_{s\bar{s}}^{i}f_s\right).
\end{equation}
Here the distribution function is normalized to the density according to $n_s(\bm{x})=\int d\bm{v} f_s$, and the Latin indices $i$ and $j$ refer to Cartesian coordinates. The collisional velocity space friction (or drag), $K_{s\bar{s}}^i$, and the collisional velocity space diffusion, $D_{s\bar{s}}^{ij}$, describe the Coulomb interaction between particles of species $s$ and $\bar{s}$.

The Coulomb friction and diffusion coefficients are compactly defined in terms of the RMJ potentials \citep{Rosenbluth_1957} 
\begin{equation}
K^i_{s\bar{s}}=-\gamma_{s\bar{s}}\frac{m_s}{m_{\bar{s}}}\frac{\partial\phi_{\bar{s}}}{\partial v^i},\quad D^{ij}_{s\bar{s}}=-\gamma_{s\bar{s}}\frac{\partial^2\psi_{\bar{s}}}{\partial v^i\partial v^j},
\end{equation}
where $\gamma_{s\bar{s}}=e_s^2e_{\bar{s}}^2\ln\Lambda/(m_s\epsilon_0)^2$. The particle phase-space potentials are weighted integrals of the distribution function
\begin{eqnarray}
\phi_{\bar{s}}(\bm{x},\bm{v})&=&-\frac{1}{4\pi}\int d\bm{\bar{v}}f_{\bar{s}}(\bm{x},\bm{\bar{v}})|\bm{v}-\bm{\bar{v}}|^{-1},\\
\psi_{\bar{s}}(\bm{x},\bm{v})&=&-\frac{1}{8\pi}\int d\bm{\bar{v}}f_{\bar{s}}(\bm{x},\bm{\bar{v}})|\bm{v}-\bm{\bar{v}}|,
\end{eqnarray}
representing the free-space Green's function solutions to the following Poisson equations
\begin{equation}
\frac{\partial^2\phi_{\bar{s}}}{\partial v^i\partial v^i}=f_{\bar{s}},\quad\frac{\partial^2\psi_{\bar{s}}}{\partial v^i\partial v^i}=\phi_{\bar{s}}.
\end{equation}
Furthermore, the diffusion and friction coefficients satisfy the relation
\begin{equation}
\frac{\partial}{\partial v^{j}}\;D^{ij}_{s\bar{s}}=\frac{m_{\bar{s}}}{m_{s}}\;K^{i}_{s\bar{s}},
\end{equation}
which can be used to show the equivalence between the Fokker-Planck operator and the Landau operator \citep{Landau_1937}
\begin{equation}
C_{s\bar{s}}[f_s,f_{\bar{s}}]=\gamma_{s\bar{s}}\frac{m_{s}}{8\pi}\frac{\partial}{\partial v^i}\int d\bm{\bar{v}}U_{ij}\left[\frac{f_{\bar{s}}}{m_s}\frac{\partial f_s}{\partial v^j}-\frac{f_s}{m_{\bar{s}}}\frac{\partial f_{\bar{s}}}{\partial \bar{v}^j}\right],
\end{equation}
where the second-rank tensor $U_{ij}$ is defined as the velocity space Hessian of the relative speed $|\bm{u}|=|\bm{v}-\bm{\bar{v}}|$ according to
\begin{equation}
U_{ij}=\frac{\partial^{2}\;u}{\partial v^i\partial v^j}=\frac{1}{u}\left(\delta^{ij}-\frac{u^iu^j}{u^2}\right).
\end{equation}
The tensor $U_{ij}$ is often called the Coulomb kernel and is a projection operator with the null-space vector $\bm{u}$. Its trace $U_{ii} = 2/u$ is the Laplacian's Green function.

\section{Poisson bracket formulation of kinetic theory}
Before working out the expression for the gyrokinetic collision operator, let us prepare the stage by introducing the use of Poisson brackets in the particle phase-space. In the paper by \citet{Brizard_2004} it was pointed out that, using the particle phase-space Poisson bracket, a momentum derivative of an arbitrary function could be written as $\{\bm{x},g\}\equiv\partial_{\bm{p}}g$. With multiple species and different masses one can apply this idea into the collision operator by following 
\begin{align}
\frac{\partial f }{\partial v^i}=m_s\{x^i,f\}_{s},
\end{align}
where the bracket $\{\cdot,\cdot\}_{s},$ refers to the bracket that is derived from the Lagrangian one-form of species $s$.

\subsection{Friction--Diffusion representation}
As the Hamiltonian equations of motion are inherently expressed with the Poisson bracket, the kinetic equation can then be formulated as \citep{Brizard_2004}
\begin{align}
\frac{\partial f_s}{\partial t}+\{f_s,H_s\}_{s}=\left\{x^i,m_s^2D_{s\bar{s}}^{ij}\{x^j,f_s\}_{s}-m_sK_{s\bar{s}}^if_s\right\}_{s},
\end{align}
where $H_s$ is the Hamiltonian for the species $s$. While in his original work Brizard used this formalism to derive a guiding-centre test particle collision operator and, later, also a bounce-averaged test-particle operator \citep[see][]{Brizard_2009}, the use of brackets was extended in the paper by \citet{Madsen_2013} to express also the friction and diffusion coefficients according to
\begin{eqnarray}
K^{i}_{s\bar{s}}&=&-\gamma_{s\bar{s}} \frac{m_s^2}{m_{\bar{s}}}\{x^i,\phi_{\bar{s}}\}_{s},\\
D^{ij}_{s\bar{s}}&=&-\gamma_{s\bar{s}} m_s^2\{x^i,\{x^j,\psi_{\bar{s}}\}_{s}\}_{s}.
\end{eqnarray}
Completing the picture for particle phase-space, we use the brackets to express also the differential equations from which the RMJ potentials are solved
\begin{equation}
m_s^2\{x^i,\{x^i,\phi_{\bar{s}}\}_{s}\}_{s}=f_{\bar{s}},\quad m_s^2\{x^i,\{x^i,\psi_{\bar{s}}\}_{s}\}_{s}=\phi_{\bar{s}}.
\end{equation}

\subsection{Landau representation}
As an alternative to the RMJ potential formulation, one may start from the Landau form of the collision operator and similarly use the bracket notation to obtain \citep[see][]{Burby_2015}
\begin{equation}
C_{s\bar{s}}[f_{s},f_{\bar{s}}]=\gamma_{s\bar{s}}\frac{m_s^2}{8\pi}\{x^i,\Gamma_{s\bar{s}}^i\}_s,
\end{equation} 
where the collisional flux $\Gamma_{s\bar{s}}^i$ is defined as the phase-space integral
\begin{equation}
\Gamma_{s\bar{s}}^i=\int d\bm{\bar{z}}\; \delta(\bm{\bar{x}}-\bm{x})U^{ij}(\bm{z},\bm{\bar{z}}) \left(f_{\bar{s}}(\bm{\bar{z}})\{x^i,f_s\}_s(\bm{z})-f_s(\bm{z})\{\bar{x}^i,f_{\bar{s}}\}_{\bar{s}}(\bm{\bar{z}})\right),
\end{equation} 
and the relative velocity needed for constructing the kernel $U^{ij}(\bm{z},\bm{\bar{z}})$ is written as
\begin{equation}
u^i(\bm{z},\bm{\bar{z}})\equiv\{x^i,H_{s}\}_{s}(\bm{z})-\{\bar{x}^i,H_{\bar{s}}\}_{\bar{s}}(\bm{\bar{z}}).
\end{equation}
Note that the brackets for both species $s$ and $\bar{s}$ are required. The delta-function in the integrand of the collisional flux $\Gamma_{s\bar{s}}^i$ represents the local nature of the collision operator in the particle phase-space but allows one to express the flux as phase-space integral rather than as a velocity-space integral. This detail is necessary for conducting the gyrocentre transformation of the Landau operator.

\subsection{Equivalence to curvilinear coordinate system}
At first sight, the idea of converting the partial derivatives to brackets might appear counterintuitive but it is in fact equivalent to expressing the derivatives in curvilinear coordinates $u^\alpha$. Since for arbitrary functions $f$ and $g$ we have according to Liouville's theorem
\begin{equation}
\{f,g\}=\frac{1}{J}\frac{\partial}{\partial u^{\alpha}}\Big[J\{f,u^{\alpha}\}g\Big],
\end{equation}
we can transform Cartesian velocity-space divergence into curvilinear coordinates according to
\begin{equation}
\frac{\partial A^i}{\partial v^i}=\frac{1}{J}\frac{\partial}{\partial u^{\alpha}}\Big[J\{x^i,u^{\alpha}\}_{s}m_{s}A^i\Big]=\frac{1}{J}\frac{\partial}{\partial u^{\alpha}}\Big[JA^{\alpha}\Big],
\end{equation}
where $J$ is the Jacobian of the transformation and $A^{\alpha}=A^i\{x^i,u^{\alpha}\}_{s}m_{s}=A^i\partial_i u^{\alpha}$ is the curvilinear component of the vector $A^i$.

This allows us to express the collision operator in a form
\begin{equation}
C[f_s,f_{\bar{s}}]=\frac{1}{J}\frac{\partial}{\partial u^{\alpha}}\Big[J\{x^i,u^{\alpha}\}_{s}m_s^2D_{s\bar{s}}^{ij}\{x^j,u^{\beta}\}_{s}\frac{\partial f}{\partial u^{\beta}}-Jm_sK_{s\bar{s}}^i\{x^i,u^{\alpha}\}_{s}f\Big].
\end{equation}
Similarly, we may express the Cartesian components of friction and diffusion coefficients according to 
\begin{eqnarray}
K_{s\bar{s}}^{i}&=&-\gamma_{s\bar{s}}\frac{m_s^2}{m_{\bar{s}}}\{x^i,u^{\beta}\}_{s}\frac{\partial\phi_{\bar{s}}}{\partial u^{\beta}},\\
D_{s\bar{s}}^{ij}&=&-\gamma_{s\bar{s}}m_s^2\Big[\{x^i,u^{\sigma}\}_{s}\{x^j,u^{\nu}\}_{s}\frac{\partial^2\psi_{\bar{s}}}{\partial u^{\sigma}\partial u^{\nu}}+\{x^i,u^{\sigma}\}_{s}\frac{\partial\{x^j,u^{\nu}\}_{s}}{\partial u^{\sigma}}\frac{\partial\psi_{\bar{s}}}{\partial u^{\nu}}\Big].
\end{eqnarray}
If we then define the transformed friction and diffusion coefficients
\begin{eqnarray}
K_{s\bar{s}}^{\alpha}&=&m_s\{x^i,u^{\alpha}\}_{s}K_{s\bar{s}}^{i},\\
D_{s\bar{s}}^{\alpha\beta}&=&m_s^2\{x^i,u^{\alpha}\}_{s}D_{s\bar{s}}^{ij}\{x^j,u^{\beta}\}_{s},
\end{eqnarray}
the collision operator can be compactly written as
\begin{equation}
C[f_s,f_{\bar{s}}]=\frac{1}{J}\frac{\partial}{\partial u^{\alpha}}\Big[JD_{s\bar{s}}^{\alpha\beta}\frac{\partial f}{\partial u^{\beta}}-JK_{s\bar{s}}^{\alpha}f\Big].
\end{equation}

The connection to curvilinear coordinates becomes transparent when we remind ourselves that the particle phase-space brackets were introduced to express velocity space derivatives and that we have the identities
\begin{eqnarray}
\{x^i,u^{\alpha}\}_{s}\{x^i,u^{\beta}\}_{s}&=&\frac{1}{m_s^2}g^{\alpha\beta},\\
\{x^i,u^{\alpha}\}_{s}\frac{\partial\{x^i,u^{\nu}\}_{s}}{\partial u^{\beta}}&=&-\frac{1}{m_s^2}g^{\alpha \ell}\Gamma_{\beta \ell}^{\nu},
\end{eqnarray}
where $g^{\alpha\beta}$ is the inverse of the metric tensor for the curvilinear velocity space coordinates $u^{\alpha}$, and $\Gamma_{\alpha\beta}^{\nu}$ are the Christoffel symbols of the second kind. Our Poisson bracket expressions for the transformed friction and diffusion coefficients are thus equivalent to
\begin{eqnarray}
K_{s\bar{s}}^{\alpha}&=&-\gamma_{s\bar{s}}\frac{m_s}{m_{\bar{s}}}\,g^{\alpha\beta}\frac{\partial\phi_{\bar{s}}}{\partial u^{\beta}},\\
D_{s\bar{s}}^{\alpha\beta}&=&-\gamma_{s\bar{s}}\Big[g^{\alpha\sigma}g^{\nu\beta}\frac{\partial^2\psi_{\bar{s}}}{\partial u^{\sigma}\partial u^{\nu}}-g^{\alpha\sigma}g^{\beta\ell}\Gamma_{\sigma\ell}^{\nu}\frac{\partial\psi_{\bar{s}}}{\partial u^{\nu}}\Big],
\end{eqnarray}
which are nothing more than the expressions one would find starting with curvilinear coordinates \citep{Goncharov_2010}.

\section{Gyrocentre collision operator}
We are finally ready to apply the gyrocentre transformation of the collision operator. It is obtained by replacing the particle phase-space Poisson bracket with the gyrocentre Poisson bracket, and by evaluating the particle phase-space quantities in terms of the gyrocentre coordinates $Z^{\alpha}$. This is the consequence of the transformation rules using the (species-wise) push-forward $\Tpush$ and pull-back $\Tpull$ operators, which can be summarized in a following manner. Scalar functions transform according to
\begin{eqnarray}
f_s(\bm{z})&=&(\Tpull F)(\bm{z})=(\Tpull F_s)(\Tpush \bm{Z})=F_s(\bm{Z}),\\
F_s(\bm{Z})&=&(\Tpush f)(\bm{Z})=(\Tpush f_s)(\Tpull \bm{z})=f_s(\bm{z}),
\end{eqnarray}
where $f_s(\bm{z})$ is understood as an arbitrary function in the particle phase space of species $s$ and $F_s(\bm{Z})$ as an image of $f_s$ at the gyrocentre phase-space. The Poisson bracket transforms under the chain rule for derivatives and the rule for scalar functions:
\begin{equation}
\{f(\bm{z}),g(\bm{z})\}=\{F(\bm{Z}),G(\bm{Z})\}_{\textrm{gy}}.
\end{equation}
Any phase-space integral transforms simply by the rule of transforming the differential volume element and the integrand giving
\begin{equation}
\int d\bm{z}f(\bm{z})\equiv\int d\bm{Z}F(\bm{Z})=\int d^6Z\mathcal{J}_{\textrm{gy}}F(\bm{Z}),
\end{equation}
where $\mathcal{J}_{\textrm{gy}}$ is the gyrocentre phase-space Jacobian.

\subsection{Friction--Diffusion representation}
Applying the transformation rules, we immediately find the gyrocentre Fokker-Planck collision operator
\begin{eqnarray}
C_{s\bar{s}}[f_s,f_{\bar{s}}]\equiv C_{s\bar{s}}^{\textrm{gy}}[F_{s},F_{\bar{s}}] & = & \left\{\Tpush X^i,\frac{}{} m_s^2\Tpush D_{s\bar{s}}^{ij}\{\Tpush X^j,F_{s}\}_{s\textrm{gy}} \right\}_{s\textrm{gy}} \nonumber \\
 &  &-\;  \left\{\Tpush X^i,\frac{}{} m_s\Tpush K_{s\bar{s}}^iF_s\right\}_{s\textrm{gy}},
\end{eqnarray}
where the expressions for the push-forwarded particle phase-space friction and diffusion coefficients are
\begin{eqnarray}
\Tpush D_{s\bar{s}}^{ij}&=&-\gamma_{s\bar{s}} m_s^2\{\Tpush X^i,\{\Tpush X^j,\Tpush \psi_{\bar{s}}\}_{s\textrm{gy}}\}_{s\textrm{gy}},\\
\Tpush K_{s\bar{s}}^{i}&=&-\gamma_{s\bar{s}} \frac{m^2_s}{m_{\bar{s}}}\{\Tpush X^i,\Tpush \phi_{\bar{s}}\}_{s\textrm{gy}}.
\end{eqnarray}
Then, using the Liouville identity, we write the gyrocentre Fokker-Planck operator in the phase-space divergence form
\begin{equation}
C_{s\bar{s}}^{\textrm{gy}}[F_s,F_{\bar{s}}]=\frac{1}{\mathcal{J}_{s\textrm{gy}}}\frac{\partial}{\partial Z^{\alpha}}\biggr[\mathcal{J}_{s\textrm{gy}}\mathcal{D}_{s\bar{s}}^{\alpha\beta}\frac{\partial F_s}{\partial Z^{\beta}}-\mathcal{J}_{s\textrm{gy}}\mathcal{K}_{s\bar{s}}^{\alpha}F_s\biggr],
\end{equation}
where the gyrocentre friction and diffusion coefficients are defined as
\begin{eqnarray}
\mathcal{D}_{s\bar{s}}^{\alpha\beta}&=&\Delta_s^{i\alpha}\;\Delta_s^{j\beta}\;\Tpush D_{s\bar{s}}^{ij},\\
\mathcal{K}_{s\bar{s}}^{\alpha}&=&\Delta_s^{i\alpha}\;\Tpush K_{s\bar{s}}^i,
\end{eqnarray}
with the so-called projection coefficients defined component-wise according to 
\begin{equation}
\Delta_s^{i\alpha}=m_s\{\Tpush X^i,Z^{\alpha}\}_{s\textrm{gy}},
\end{equation}
and being analogous to Cartesian components of contravariant basis vectors.

Using the projection coefficients and the chain rule for the Poisson brackets, the expressions for the push-forwarded particle phase-space friction and diffusion coefficients become
\begin{eqnarray}
\Tpush D_{s\bar{s}}^{ij}&=&-\gamma_{s\bar{s}} \Delta_s^{i\sigma}\frac{\partial}{\partial Z^{\sigma}}\left(\Delta_s^{j\nu}\frac{\partial}{\partial Z^{\nu}}\Tpush \psi_{\bar{s}}\right),\\
\Tpush K_{s\bar{s}}^{i}&=&-\gamma_{s\bar{s}}\frac{m_s}{m_{\bar{s}}}\Delta_s^{i\sigma}\frac{\partial}{\partial Z^{\sigma}}\Tpush \phi_{\bar{s}}.
\end{eqnarray}
After introducing the symmetric matrix 
\begin{equation}
\Xi_s^{\alpha\beta}=\Delta_s^{i\alpha}\Delta_s^{i\beta},
\end{equation}
we can finally express the gyrocentre friction and diffusion coefficients according to
\begin{eqnarray}
\label{eq:Ka}
\mathcal{K}_{s\bar{s}}^{\alpha}&=&-\gamma_{s\bar{s}}\frac{m_s}{m_{\bar{s}}}\;\Xi_s^{\alpha\beta}\; \frac{\partial\Tpush \phi_{\bar{s}}}{\partial Z^{\beta}},\\
\label{eq:Dab}
\mathcal{D}_{s\bar{s}}^{\alpha\beta}&=&-\frac{\gamma_{s\bar{s}} }{2}\biggr[\left(\Xi_s^{\alpha\sigma}\Xi_s^{\beta\nu}+\Xi_s^{\beta\sigma}\Xi_s^{\alpha\nu}\right)\frac{\partial^2\Tpush \psi_{\bar{s}}}{\partial Z^{\sigma}\partial Z^{\nu}}\nonumber\\ & &\qquad\qquad\qquad+\left(\Xi_s^{\alpha\sigma}\Delta_s^{k\beta}+\Xi_s^{\beta\sigma}\Delta_s^{k\alpha}\right)\frac{\partial\Delta_s^{k\nu}}{\partial Z^{\sigma}}\frac{\partial\Tpush \psi_{\bar{s}}}{\partial Z^{\nu}} \biggr],
\end{eqnarray}
where the symmetry in $\mathcal{D}_{s\bar{s}}^{\alpha\beta}$ with respect to $\alpha$ and $\beta$ has been built-in using the fact that the particle phase-space diffusion coefficient is symmetric with respect to $i$ and $j$. 

Referring to our earlier discussion of the particle phase-space operator in curvilinear coordinates, we see that the gyrocentre friction and diffusion coefficients appear no more exotic: they essentially have the same form as the particle phase-space friction and diffusion coefficients in curvilinear coordinates.

\subsection{Gyrocentre Rosenbluth-MacDonald-Judd potential equations}
To complete the gyrocentre transformation of the Fokker-Planck operator, expressions for the gyrocentre RMJ potentials are needed. Two different approaches are possible. One can obtain the potentials directly by transforming the particle phase-space integral definitions into the gyrocentre phase-space according to 
\begin{eqnarray}
\label{eq:integral_gy_phi}
\Tpush \phi_{\bar{s}}&=&-\frac{1}{4\pi}\int d\bm{\bar{Z}}\;\delta^{\textrm{gy}}\frac{F_{\bar{s}}}{u^{\textrm{gy}}},\\
\label{eq:integral_gy_psi}
\Tpush \psi_{\bar{s}}&=&-\frac{1}{8\pi}\int d\bm{\bar{Z}}\;\delta^{\textrm{gy}}F_{\bar{s}}\;u^{\textrm{gy}},
\end{eqnarray}
where the definitions for the gyrocentre delta-function and relative velocity are
\begin{eqnarray}
\label{eq:gy_delta}
\delta^{\textrm{gy}}(\bm{Z},\bm{\bar{Z}})&=&\delta(\Tpush\bm{X}-\Tpushbar\bm{\bar{X}}),\\
\label{eq:gy_u}
\bm{u}^{\textrm{gy}}&=&\{\Tpush\bm{X},{\cal H}_{s}\}_{s}-\{\Tpushbar\bm{\bar{X}},{\cal H}_{\bar{s}}\}_{\bar{s}},
\end{eqnarray}
and ${\cal H}$ is the gyrocentre Hamiltonian. Alternatively, one may push the particle phase-space Poisson equations for the RMJ potentials into the gyrocentre phase-space according to 
\begin{eqnarray}
m_s^2\{\Tpush X^i,\{\Tpush X^i,\Tpush \phi_{\bar{s}}\}_{s\textrm{gy}}\}_{s\textrm{gy}}&=&F_{\bar{s}},\\
m_s^2\{\Tpush X^i,\{\Tpush X^i,\Tpush \psi_{\bar{s}}\}_{s\textrm{gy}}\}_{s\textrm{gy}}&=&\Tpush \phi_{\bar{s}}.
\end{eqnarray}
Converting the double brackets into double phase-space divergences similarly as was done for the collision operator, we find phase-space Poisson-like equations for the gyrocentre RMJ potentials
\begin{eqnarray}
\label{eq:diff_gy_phi}
\frac{1}{\mathcal{J}_{s\textrm{gy}}}\frac{\partial}{\partial Z^{\alpha}}\biggr(\mathcal{J}_{s\textrm{gy}}\Xi_s^{\alpha\beta}\frac{\partial \Tpush \phi_{\bar{s}}}{\partial Z^{\beta}}\biggr)&=&F_{\bar{s}},\\
\label{eq:diff_gy_psi}
\frac{1}{\mathcal{J}_{s\textrm{gy}}}\frac{\partial}{\partial Z^{\alpha}}\biggr(\mathcal{J}_{s\textrm{gy}}\Xi_s^{\alpha\beta}\frac{\partial \Tpush \psi_{\bar{s}}}{\partial Z^{\beta}}\biggr)&=&\Tpush \phi_{\bar{s}},
\end{eqnarray}
where the matrix $\Xi_s^{\alpha\beta}$ now clearly is analogous to the metric in curvilinear coordinates.

\subsection{Landau representation}
Similarly, using the transformation rules, one may compute the gyrocentre transformation of the particle phase-space Landau operator \citep[for details, see][]{Burby_2015}. When written in the friction-diffusion form, the coefficients in the gyrokinetic Landau operator become
\begin{eqnarray}
\mathcal{D}_{s\bar{s}}^{\alpha\beta}(\bm{Z})&=&\frac{\gamma_{s\bar{s}}}{8\pi}\Delta_s^{i\alpha}\Delta_s^{j\beta}\int d\bm{\bar{Z}}\delta^{\textrm{gy}}U^{ij}F_{\bar{s}},\\
\mathcal{K}_{s\bar{s}}^{\alpha}(\bm{Z})&=&\frac{\gamma_{s\bar{s}}}{8\pi}\frac{m_s}{m_{\bar{s}}}\Delta_s^{i\alpha}\int d\bm{\bar{Z}}\delta^{\textrm{gy}}U^{ij}\bar{\Delta}_{\bar{s}}^{j\beta}\frac{\partial F_{\bar{s}}}{\partial \bar{Z}^{\beta}},
\end{eqnarray}
with $U^{ij}$ now a function of the gyrocentre relative velocity $\bm{u}^{\textrm{gy}}$ defined in Eq.~(\ref{eq:gy_u}). As such, these coefficients would seem to be simpler than the expressions we derived in Eqs.~(\ref{eq:Ka}) and~(\ref{eq:Dab}) but they are difficult to evaluate due to the complicated expression for the gyrocentre delta-function $\delta^{\textrm{gy}}(\bm{Z},\bm{\bar{Z}})$ defined in Eq.~(\ref{eq:gy_delta}). 

\section{Unresolved aspects of the gyrokinetic collision operator}
As the ultimate goal of gyrokinetic theory is to eliminate the fast gyromotion time scale, the gyrocentre collision operator must be gyroaveraged. Assuming that the distribution functions are gyroangle independent
\citep[for the ordering, see][]{Brizard_2004}, the gyroaveraged 5D collision operator is written as
\begin{equation}
\label{eq:gyroaveraged_collision_operator}
\left\langle C_{s\bar{s}}^{\textrm{gy}}[\left\langle F_s\right\rangle,\left\langle F_{\bar{s}}\right\rangle]\right\rangle=\frac{1}{\mathcal{J}_{s\textrm{gy}}}\frac{\partial}{\partial Z^{\alpha}}\biggr[\mathcal{J}_{s\textrm{gy}}\left\langle\mathcal{D}_{s\bar{s}}^{\alpha\beta}\right\rangle\frac{\partial \left\langle F_s\right\rangle}{\partial Z^{\beta}}-\mathcal{J}_{s\textrm{gy}}\left\langle\mathcal{K}_{s\bar{s}}^{\alpha}\right\rangle\left\langle F_s\right\rangle\biggr].
\end{equation}
The important question is what happens for the gyrokinetic RMJ potentials and the corresponding differential equations, i.e., can the gyrocentre potential equations be reduced to 5D or not.

\subsection{Existence of 5D potential functions and Green's function operators?}
In particle phase-space, a distribution function with axial symmetry in the velocity space provides particle phase-space RMJ potentials that have the same axial symmetry. In gyrocentre phase-space, only in the limit of zero Larmor radius have we been able to explicitly verify that the potentials indeed become gyroangle independent. At the zero-Larmor-radius limit, both the gyroaveraged gyrokinetic Landau collision integral and the differential equations defining the gyrokinetic RMJ potentials reduce to the corresponding axially symmetric particle phase space expressions due to the gyrocentre transformation collapsing to an identity. If the FLR effects are included, the gyrokinetic RMJ potentials remain gyroangle dependent. This is reflected in the Eqs.~(\ref{eq:diff_gy_phi}) and~(\ref{eq:diff_gy_psi}) in the sense that the tensor $\Xi_s^{\alpha\beta}$ is gyroangle dependent whereas, in the particle phase-space, the velocity-space metric tensor, in either cylindrical or spherical coordinates, is axially symmetric. In short: given gyroangle independent gyrocentre distribution function, the gyrocentre potentials cannot be assumed gyroangle independent, and the differential equations~(\ref{eq:diff_gy_phi}) and~(\ref{eq:diff_gy_psi}) cannot be reduced to 5D. 

One could, of course, assume the gyroangle dependency of the gyrokinetic potentials to be weak and approximate
\begin{align}
\Tpush \phi_{\bar{s}}&\approx \left\langle\Tpush \phi_{\bar{s}}\right\rangle,\\
\Tpush \psi_{\bar{s}}&\approx \left\langle\Tpush \psi_{\bar{s}}\right\rangle.
\end{align}
This would allow one to average the differential
equations~(\ref{eq:diff_gy_phi}) and~(\ref{eq:diff_gy_psi}) over the
gyroangle and to reduce them to 5D. This approximation will, however,
most likely lead to violation of the important conservation properties
of the gyrokinetic collision operator. Indeed, the latter can be
proven only in the Landau formulation and the Landau and
potential formulations are equivalent only if the gyroangle dependency
of the potentials is preserved.

From a theoretical point-of-view, it would also be aesthetic to explicitly verify if the differential operator we have derived for the gyrocentre RMJ potentials in Eqs. (\ref{eq:diff_gy_phi}) and (\ref{eq:diff_gy_psi}) satisfies
\begin{eqnarray}
\frac{1}{\mathcal{J}_{s\textrm{gy}}}\frac{\partial}{\partial Z^{\alpha}}\biggr[\mathcal{J}_{s\textrm{gy}}\Xi_s^{\alpha\beta}\frac{\partial}{\partial Z^{\beta}}\left( \frac{\delta^{\textrm{gy}}}{u^{\textrm{gy}}}\right)\biggr] &=&-4\pi \delta(\bm{Z}-\bar{\bm{Z}}),\\
\frac{1}{\mathcal{J}_{s\textrm{gy}}}\frac{\partial}{\partial Z^{\alpha}}\biggr[\mathcal{J}_{s\textrm{gy}}\Xi_s^{\alpha\beta}\frac{\partial}{\partial Z^{\beta}}\Big( \delta^{\textrm{gy}}u^{\textrm{gy}}\Big)\biggr] &=& 2\frac{\delta^{\textrm{gy}}}{u^{\textrm{gy}}}.
\end{eqnarray}
This is a property one would expect given the integral presentations (\ref{eq:integral_gy_phi}) and (\ref{eq:integral_gy_psi}).

\subsection{Numerical considerations}
Regardless of whether one opts for the Landau formulation or for the
potential formulation, the computational effort in evaluating the
gyrokinetic collision operator accurately will be
significant. As an alternative, a stochastic approach could be chosen for
emulating the Fokker-Planck operator using the corresponding
stochastic differential equations \citep[for a discussion on guiding-centre test-particle operator, see][]{Hirvijoki_2013} which require only the evaluation of the friction and diffusion
coefficients. The stochastic approach is, however, a completely different paradigm that is set aside in the present discussion. 

Choosing the Landau formulation requires evaluation of a 6D integral over phase-space, regardless whether the distribution functions are assumed gyroangle independent. The reason is that the function $\delta^{\textrm{gy}}(\bm{Z},\bm{\bar{Z}})$ is implicitely gyroangle dependent on both coordinate spaces and finding the many zeros of its argument is a difficult (if not intractable) non-linear problem. Choosing the formulation in terms of the gyrokinetic RMJ potentials, 6D Poisson-like differential equations need to be inverted instead. Evaluating weighted integrals for $n$ points would result in a minimum effort of the order ${\cal O}(n^2)$ whereas there is evidence that elliptic equations can be inverted with ${\cal O}(n\log n)$ effort \citep[see][]{Pataki_2011}. As the number of points $n$ at which the distribution function is presented can be expected to be large, the potential formulation could thus offer significant speed-up for the evaluation of the gyrokinetic collision operator, even if the boundary conditions for the potential equations must be computed from the integral expressions (there typically are far less boundary points than the total number of points).

As a final difficulty, the discrete implementation would have to satisfy the same conservation properties as the continuous operator does. In the case of axially symmetric particle phase-space operator or, equivalently, in the zero-Larmor-radius limit of the gyrokinetic operator, conservative numerical methods have been developed for both the Landau and the potential formulations (see e.g. \citet{Hager_2016,Taitano_2015}). On the other hand, it has been explicitly shown in \citet{Burby_2015} that energetically consistent collisional gyrokinetics requires both the Vlasov and the collision operator to be treated equally at the same order with respect to the asymptotic gyrocentre transformation. Whether a conservative numerical method can be found for the gyrocentre operator including the FLR effects remains to be seen.

\section{Summary}
In this paper, a differential formulation for the Lie-transformed gyrocentre collision operator has been derived. This was achieved by transforming not only the particle phase-space Fokker-Planck friction and diffusion coefficients but also the particle phase-space Poisson equations that determine the Rosenbluth-MacDonald-Judd potential functions needed in evaluating the friction and diffusion coefficients. Our final results are summarized in the expressions for the gyroaveraged collision operator defined in Eq.~(\ref{eq:gyroaveraged_collision_operator}), the gyrocentre friction and diffusion coefficients defined in Eqs.~(\ref{eq:Ka}) and~(\ref{eq:Dab}), and the gyrocentre equivalents for the Rosenbluth-MacDonald-Judd potential functions defined either through the integrals~(\ref{eq:integral_gy_phi}) and~(\ref{eq:integral_gy_psi}) or the differential equations~(\ref{eq:diff_gy_phi}) and~(\ref{eq:diff_gy_psi}).

We also argued that the gyrocentre Rosenbluth-MacDonald-Judd potentials are not expected to be gyroangle independent, even if the gyrocentre distribution functions were to be. Thus, the differential equations for the gyrocentre Rosenbluth-MacDonald-Judd potentials cannot be averaged over the gyroangle but remain 6D. However, with both integral and differential expressions available for the gyrocentre potential functions, we expect that fast elliptic solvers could be applied to solve the differential equations with boundary conditions evaluated from the integral expressions, similarly as is commonly done for the particle phase-space Rosenbluth-MacDonald-Judd potentials. As such, this recipe could offer significant speed-up from ${\cal O}(n^2)$ for the pure integral definitions to ${\cal O}(n \log n)$ if the integral definitions are used only for evaluating boundary conditions for the differential equations.

\begin{acknowledgements}
The authors are grateful for the discussions with Hong Qin, Joshua Burby, Manasvi Lingam, Amitava Bhattacharjee, and the anonymous Referees. The work of EH and DP is supported by the Department of Energy Contract No. DE-AC02-09CH11466 and the work of AJB is supported by the Department of Energy grant No. DE-SC0014032.
\end{acknowledgements}

\bibliographystyle{jpp}

\bibliography{final}

\end{document}